\documentclass[prl,superscriptaddress,showpacs,twocolumn]{revtex4-1}

\usepackage{graphicx}
\usepackage{color}
\usepackage{graphicx}
\usepackage{amssymb}
\usepackage{epstopdf}
\usepackage{times}
\usepackage{ulem}
\begin{document}


\title{Small scale rotational disorder observed in epitaxial graphene on SiC(0001)} 
\date{\today}


\author{Andrew L. Walter$^{1,2,3,\footnotemark}$, Aaron Bostwick$^{1}$, Florian Speck$^{4}$, Markus Ostler$^{4}$, Keun Su Kim$^{1}$, Young Jun Chang$^{1,2}$, Luca Moreschini$^{1}$,Davide Innocenti$^{5}$,Thomas Seyller$^{4}$, Karsten Horn$^{2}$, Eli Rotenberg}

\address{ Advanced Light Source (ALS), E. O. Lawrence Berkeley National Laboratory, Berkeley, California 94720, USA.\\ $^{2}$ Department of Physical Chemistry, Fritz-Haber-Institut der Max-Planck-Gesellschaft, Faradayweg 4-6, 14195 Berlin, Germany.\\$^{3}$Donostia International Physics Centre, Paseo Manuel de Lardizabal, 4. 20018Donostia- San Sebastian, Spain  \\$^{4}$ Lehrstuhl f\"{u}r Technische Physik, Universit\"{a}t Erlangen-N\"{u}rnberg, Erwin- Rommel-Strasse 1, 91058 Erlangen, Germany.\\$^{5}$University of Rome (Tor Vergata), Rome 00173, Italy} 

\footnotetext{corresponding author: andrew_walter@ehu.es}



\date{\today}

\def\EF{$E_{\mathrm{F}}$}
\def\ED{$E_{\mathrm{D}}$}
\def\kk{$\mathrm{\overline{KK}}$}
\def\gk{$\mathrm{\overline{\Gamma K}}$}
\def\gkm{$\mathrm{\overline{\Gamma K M}}$}
\def\buf{$6\sqrt{3}$}
\definecolor{eli}{rgb}{0,0,0}
\def\eli{\textcolor{eli}}
\definecolor{Andrew}{rgb}{1,0,0}
\def\Andrew{\textcolor{Andrew}}
\definecolor{Thomas}{rgb}{0,1,0}
\def\Thomas{\textcolor{Thomas}}

\begin{abstract}
Interest in the use of graphene in electronic devices has motivated an explosion in the study of this remarkable material. The simple, linear Dirac cone band structure offers a unique possibility to investigate its finer details by angle-resolved photoelectron spectroscopy (ARPES). Indeed, ARPES has been performed on graphene grown on metal substrates but electronic applications require an insulating substrate. Epitaxial graphene grown by the thermal decomposition of silicon carbide (SiC) is an ideal candidate for this due to the large scale, uniform graphene layers produced. The experimental spectral function of epitaxial graphene on SiC has been extensively studied. However, until now the cause of an anisotropy in the spectral width of the Fermi surface has not been determined. In the current work we show, by comparison of the spectral function to a semi-empirical model, that the anisotropy is due to small scale rotational disorder ($\sim\pm$ 0.15$^{\circ}$) of graphene domains in graphene grown on SiC(0001) samples. In addition to the direct benefit in the understanding of graphene's electronic structure this work suggests a mechanism to explain similar variations in related ARPES data.
\end{abstract}


\maketitle 


Interest in the single layer of hexagonally coordinated carbon atoms, known as graphene, has been intense ever since the discovery of its unusual electronic properties \cite{Geim:2007p4020}. An understanding of the electronic properties is essential if graphene applications are to be realised. Angle Resolved Photoemission Spectroscopy (ARPES) measurements provide the most direct method to investigate the electronic band structure, however this technique requires large well defined samples. ARPES have been performed on graphene on metal substrates, however electronic applications will require insulating or semiconducting substrates. A significant number of ARPES studies have therefore been performed on graphene grown epitaxially on SiC by thermal decomposition\cite{Berger:2004jh,Emtsev:2009p212}, a preparation method which is a very promising candidate for applications because the production process can be scaled up, uses common semiconductor processing steps and provides large scale uniform layers. 

\begin{figure}[b]
\includegraphics{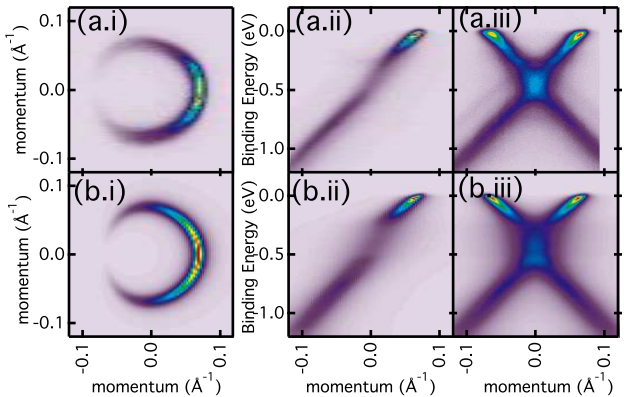}
\caption{\label{spectral}Experimental (a) and semi- empirical model (b) Fermi surfaces (i) and spectral functions in the \gk\ (ii) and \kk\ (iii) directions of epitaxial graphene on SiC(0001).}
\end{figure}

This material can be considered an important testbed for exciting solid state physics, and interest has turned to fine details of its band structure, as observed by ARPES. In this context we compare ARPES measurements to a semi-empirical photoemission model. Comparison of the model to experimental data accounted for almost all features of the electronic structure, with the exception of an anisotropy in the spectral width. We show that this anisotropy is explained by a $\pm$ 0.15$^{\circ}$ rotational disorder of the graphene domains. Fig. \ref{spectral} (a) shows the well known experimental spectral function obtained from epitaxial graphene on SiC(0001), \ED\ $\sim$ -0.5 eV,  which was demonstrated to sit on top of the so-called buffer layer with $(6\sqrt{3} \times 6\sqrt{3}) R 30^{\circ}$ periodicity (hence forth \buf) \cite{Emtsev:2008p4975}. The anistropy in the Fermi surface intensity \cite{MuchaKruczynski:2008p214}, the increase in the intensity between the Fermi surface and $\sim$ 200 meV \cite{McChesney:2008p261} and the offset of the bands above and below the Dirac crossing, \ED\ , \cite{Polini:2008p3589,Hwang:2008p3441,Bostwick:2010p3387} have all been successfully described. Until now the anisotropy in the spectral width of the Fermi surface seen in Fig. \ref{spectral} (a.i) as an increased width in the vertical, $k_{y}$ (\kk), direction when compared to that in the horizontal, $k_{x}$ (\gk), direction has not been explained. 

This variation is not confined to the Fermi surface, but appears across all energies. This is important for a detailed analysis of the spectral function in many contexts such as gap formation, the examination of many-body interactions and the investigation of the the Berry's phase in graphene. The current work provides a simple explanation of this feature, in terms of small scale rotational disorder, which is essential to the discussion of the electronic structure of graphene. In fact many papers have used intensity and line-width variations in ARPES data to discuss the existence of various quasiparticles\cite{Qi:2010p4916,Gierz:2011do,Siegel:2011va,Kim:2008p4920,Zhou:2007p4898,Bostwick:2010p3387,Bostwick:2009p256} and in discussions of the Berry's phase \cite{Gierz:2011do,deGail:2011cl,Bostwick:2007p252} in graphene. The current work therefore provides an invaluable insight into ARPES spectral variations and in the understanding of these more exotic features.

\begin{figure}[b]
\includegraphics{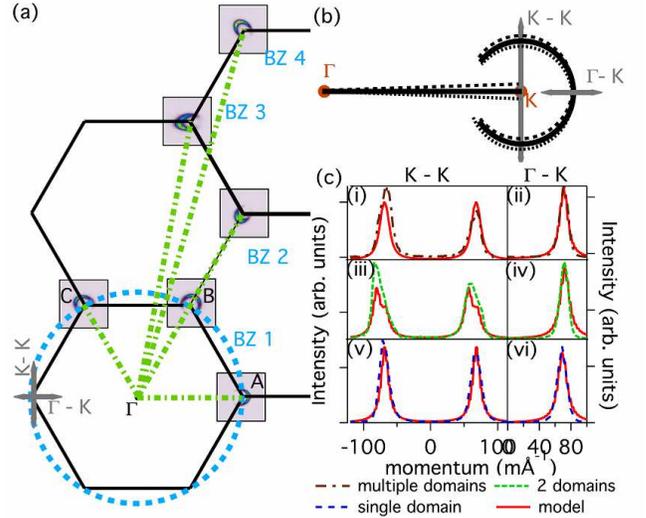}
\caption{\label{diagram}Diagram showing the (a) Brillouin zone boundaries for graphene with the 1st, 2nd, 3rd and 4th Brillouin zone K point radial lines (dashed, blue, lines) of graphene, (b) schematic of a selected K point showing the effect of small scale rotational disorder and (c) selected experimental and semi- empirical model intensity profiles in the \kk\ and \gk\ directions. In (a) the Fermi surfaces indicate the K points at which measurements where undertaken.}
\end{figure}

\begin{figure*}
\includegraphics{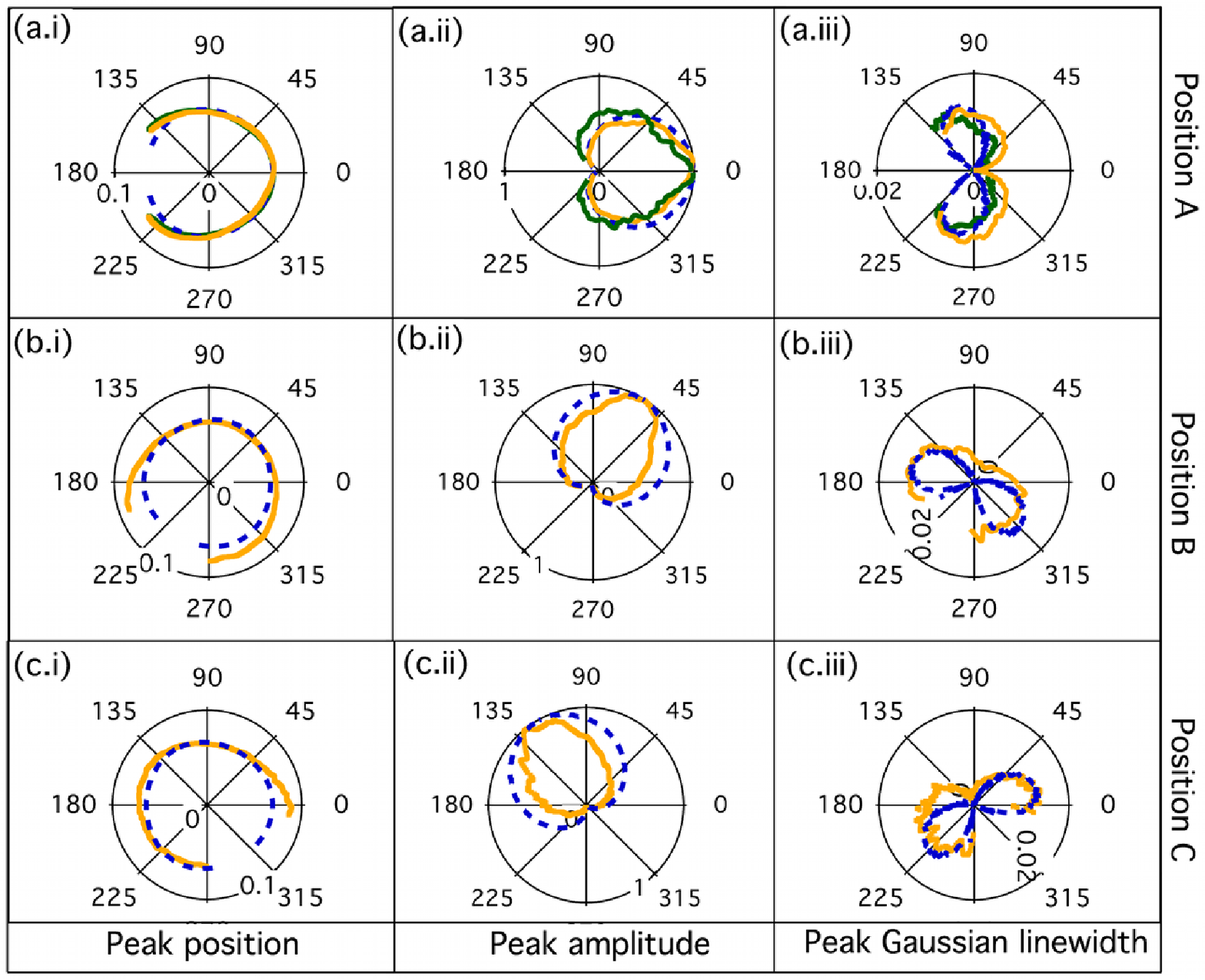}
\caption{\label{zone1}Fitted Fermi surface data for vacuum grown (solid light, orange, lines), argon grown (solid dark, green, lines) and semi- emperical model (dashed, blue, lines) for the three 1st Brillouin K points indicated in Fig. \ref{diagram}. Position(i), amplitude(ii) and Gaussian Linewidths (iii) obtained from Lorentzian-Gaussian fits to the Fermi surface are presented. In all cases the Lorentzian width is set to 0.15 $\AA^{-1}$ ensuring that the Gaussian width describes the variation in the width of the Fermi surface.}
\end{figure*}

\begin{figure*}
\includegraphics{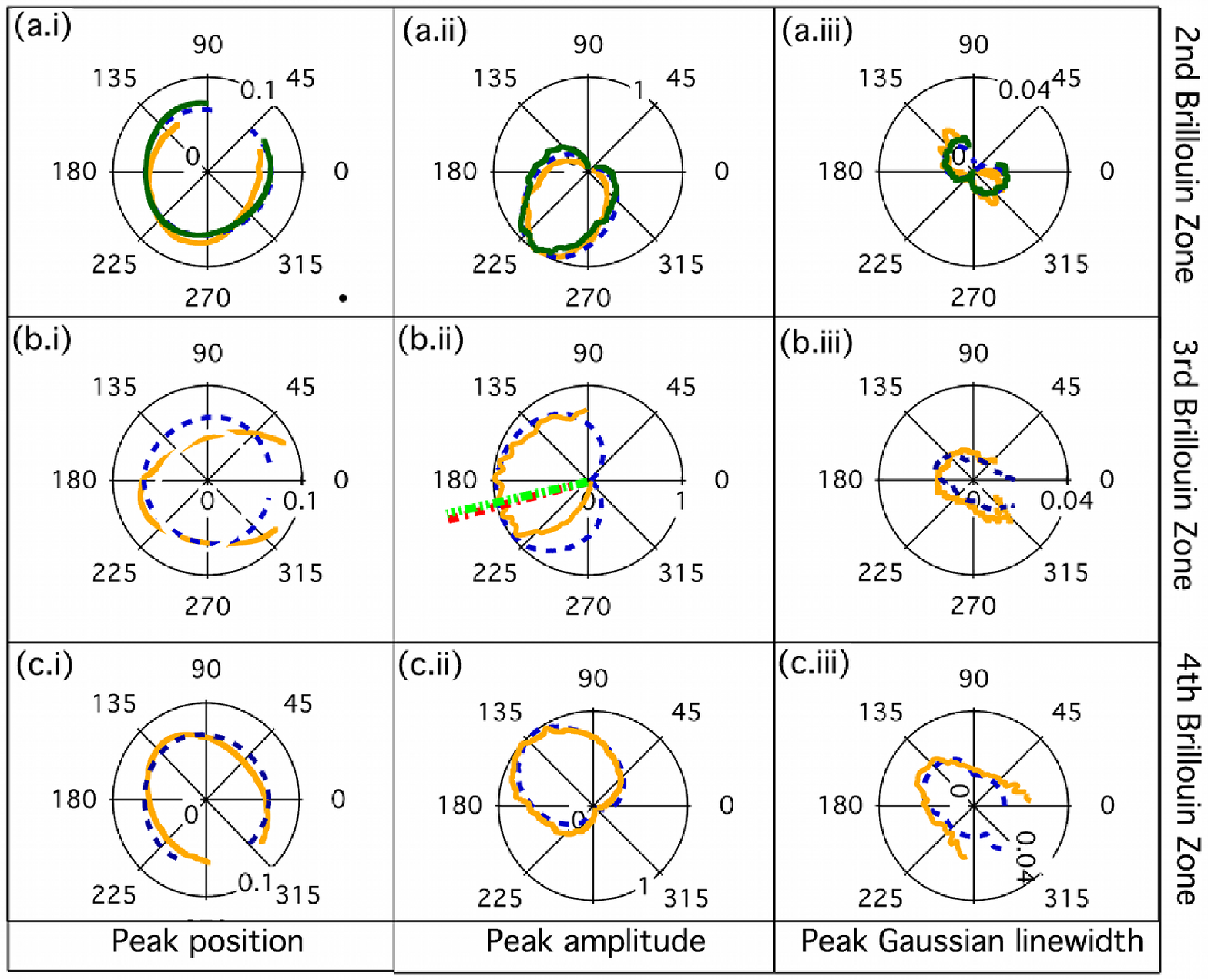}
\caption{\label{higherzone}Fitted Fermi surface data for vacuum grown (solid light, orange, lines), argon grown (solid dark, green, lines) and semi emperical model (dashed, blue, lines) for the higher order Brillouin zone K points indicated in Fig. \ref{diagram}. Position (i), amplitude (ii) and Gaussian Linewidths (iii) obtained from Lorentzian-Gaussian fits to the Fermi surface are presented. In all cases the Lorentzian width is set to 0.15 $\AA^{-1}$ ensuring that the Gaussian width describes the variation in the width of the Fermi surface. The red and green lines in b ii indicate the position of the maximum intensity.}
\end{figure*}

ARPES were obtained at the Maestro end station (SES-R4000 analyzer) at beamline 7 of the Advanced Light Source, Lawrence Berkeley National Laboratory. Spectra where taken at $\sim$ 20 K, at $<$  2x10$^{-10}$ Torr using 95 eV photons giving an overall resolution of $\sim$ 25 meV and 0.01 \AA$^{-1}$. Epitaxial graphene on SiC(0001) was prepared either by annealing in ultra-high vacuum (UHV)\cite{Berger:2004jh} (vacuum grown) or by annealing in Ar \cite{Emtsev:2009p212} (argon grown) using a custom build reactor\cite{Ostler:2010p5375}. The decomposition of SiC in Ar leads to an improved crystalline quality \cite{Emtsev:2009p212}.

The bare band used in the semi-empirical ARPES model is a first nearest neighbour tight binding (FNN TB) model fit to experimental graphene on \buf\ data. This models the $\pi$ bands via the relations:

\begin{equation}
\mathrm{E_{bare}(\bf k \rm) = \frac{\epsilon_{2p}\pm \gamma_0\omega(\bf k\rm)}{(1 \pm s_0\omega(\bf k \rm)}}
\end{equation}
\begin{equation}
\mathrm{\omega(\bf k \rm) = \sqrt{1+4cos(\frac{\sqrt{3}ak_y}{2})cos(\frac{ak_x}{2}) + 4cos^2(\frac{ak_x}{2})}}
\end{equation}

with a lattice constant, a = 2.46 \AA, and the fitted parameters, $\gamma_{0}$=-3.24 eV and s$_0$=0.0425 eV, are those found by Bostwick et al.\cite{Bostwick:2007p252}. The final fitted parameter, $\epsilon_{2p}$, is the offset of the Dirac energy, \ED\ , from the Fermi level due to doping of the graphene by the substrate and is determined by comparison to the ARPES measurements. Broadening of the bare band is introduced to the model through the self energy via the spectral function relation\cite{Bostwick:2009p255}

\begin{equation}
\mathrm{ \bf A\rm (E, \bf k \rm) = \frac{|Im \Sigma(E, \bf k \rm)|}{(E - E_{bare}(\bf k\rm) - Re\Sigma(E, \bf k \rm))^2 + Im\Sigma(E, \bf k \rm)^2} }
\end{equation}
 
The self energy is determined using the semi-empirical method of Bostwick et al. \cite{Bostwick:2007p247}, where the linewidth of the ARPES momentum density curves (MDCs) are used to determine the imaginary component of the self energy, which is Hilbert transformed to get the real component. This experimental self energy is then used to recreate the experimental data and a self-consistent fitting used to further refine the self energy. An additional Gaussian broadening term is added to the current model to account for experimental broadening, $dE = 25 meV$.

Rotational disorder of graphene on metal substrates such as Cu(111)\cite{Walter:2011uu} and Pt(111)\cite{Sutter:2009ff} has been shown to lead to significant anisotropic spectral broadening, therefore we attempted to model the much smaller anisotropic broadening seen in Graphene on SiC with a similar small rotational disorder. To model the effect of rotational disorder several 3D spectral functions, rotated around the $\Gamma$ point by small angles, are summed together. This is illustrated, for a 2D Fermi surface, in Fig. \ref{diagram} (b). The non-rotated Fermi surface is the solid arc around the K point, two further Fermi surfaces (dashed arcs) are shown rotated by a small amount around $\Gamma$. The effect on the Fermi surface is that it appears broader in the \kk\ direction than in the \gk\ direction. Examples of the Fermi surface intensity profiles in the \kk\ and \gk\ directions for three different regions on a sample are shown in Fig. \ref{diagram} (c). Fitted Fermi surface data presented in Fig. \ref{zone1} and Fig. \ref{higherzone} are obtained by taking similar intensity line profiles of the Fermi surface in 1 degree steps around the K point with the \gk\ direction being set as the zero angle. The Lorentzian width is set to the width of the Fermi surface in the \gk\ direction (0.15 \AA$^{-1}$), ensuring the Gaussian line width describes the variation in width of the Fermi surface. 

In order to investigate the anisotropy in the width of the Fermi surface, experimental spectral functions at a number of K points in the graphene Brillouin zone ( see Fig. \ref{diagram} (a)) were obtained. Fitted Fermi surface data were then extracted from these spectral functions and are presented in Fig. \ref{zone1} and Fig. \ref{higherzone}. Spectral functions obtained using the semi-empirical model described in the methods section were also analysed in a similar manner and are overlayed on the data in Fig. \ref{zone1} and Fig. \ref{higherzone}.

The experimental data obtained from the first Brillouin zone (position A, B and C from Fig. \ref{diagram}) is presented in Fig. \ref{zone1}. This data is compared to a semi- empirical model which involves 15 rotational domains equally spaced between $\pm 0.15^{\circ}$. The size (peak position) and intensity (peak amplitude) of the Fermi surface is well described by the model. The anisotropy in the width of the Fermi surface (variation in the peak Gaussian linewidth) is clearly evident by the "peanut" shape in Fig. \ref{zone1} (a.iii), (b.iii) and (c.iii) for both vacuum grown and argon grown samples. The "peanut" shape is only produced in the semi-empirical model by considering rotational disorder. The size of the "peanut" lobes is directly proportional to the rotational variation of the domains, with $\pm 0.15^{\circ}$ providing the best agreement.

Due to the six-fold symmetry the first Brillouin zone K points are all equivalent with respect to the proposed rotational disorder. This is not the case for the higher order Brillouin zones (BZ2, BZ3 and BZ4 in Fig. \ref{diagram}) where the axis of rotation no longer lies in the \gk\ direction. Investigation of the higher Brillouin zone experimental spectral functions (presented in Fig. \ref{higherzone}) then provides an important test of the rotational disorder model. The size (peak position) and intensity (peak amplitude) of the Fermi surfaces at each of the higher Brillouin zone locations are again well described by the rotational disorder model (15 domains, $\pm 0.15^{\circ}$ rotational spread). In particular Fig. \ref{higherzone} (b.ii) indicates that the experimental Fermi surface intensity has a maximum at $\sim$ 190$^{\circ}$ (dash dot dot, light green, line) rather than the expected 180$^{\circ}$, which is predicted by the rotational disorder model (dash dot, red, line).

The "peanut" shape of the Fermi surface anisotropy (Fig. \ref{higherzone} (a.iii) ) is observed in the second Brillouin zone, However this becomes more oval shaped in the third (b.iii) and fourth (c.iii) Brillouin zones as the axis of rotational disorder no longer aligns with the intensity anisotropy axis \cite{MuchaKruczynski:2008p214}. This variation is also well described by the rotational disorder model. 

Further evidence for the rotational disorder is found in experimental Fermi surface intensity profiles (dashed lines) in the \kk\ and the \gk\ directions (Fig. \ref{diagram} (c) taken from three distinct regions on an argon grown sample.The intensity profile from the semi- empirical model, with the correct rotational disorder, is overlayed in red (solid line). The spectra were obtained from the first Brillouin zone K point and show a similar width and shape in the \gk\ direction (Fig. \ref{diagram} (c.ii),(c.iv) and (c.vi)). In contrast three distinct line shapes are observed in the \kk\ direction (Fig. \ref{diagram} (c.i),(c.iii) and (c.v)).

The most commonly observed lineshape across several samples is shown in Fig. \ref{diagram} (c.i) and is well described by the rotational disorder model (15 domains within the $\pm 0.15^{\circ}$ rotational spread).The other two spectra are rare compared to the rotationally disordered spectra but provide a significant insight. In Fig. \ref{diagram} (c.iii) an asymmetric line shape is observed, which is described by a model including only 2 rotational domains ($\pm 0.15^{\circ}$) with an intensity ratio of 2:1 between them. The final region (Fig. \ref{diagram} (c.v)) has a similar lineshape in the \kk\ and \gk\ direction, corresponding to a region containing only a single rotation.

Comparison of experimental Angle Resolved Photoemission Spectra from "vacuum grown" and "argon grown" epitaxial graphene on SiC(0001) to a semi-empirical model confirms the existence of a $\le \pm 0.15^{\circ}$ rotational disorder of the Fermi surface. This disorder is attributed to a number of rotated graphene domains within the 50$\mu$m photon beam size. Experimental data from regions of a vacuum grown sample which show only 1 and 2 rotational domains are also presented, however these are rare compared to the rotationally disordered spectra. Importantly the scale of the rotation ($\pm 0.15^{\circ}$) is much smaller than is possible to determine from modern imaging techniques, like LEEM/PEEM and STM, and the common diffraction techniques, LEED and RHEED. It is therefore shown that detailed analysis of ARPES features can give information on small scale structure variations not possible until now.

\begin{acknowledgments}
The Advanced Light Source is supported by the Director, Office of Science, Office of Basic Energy Sciences, of the U.S. Department of Energy under Contract No. DE-AC02-05CH11231. Work in Erlangen was supported by the ESF program EuroGRAPHENE and by the DFG priority program 1459 \textit{Graphene}. A.W. acknowledges support form the Max Planck Society.K.S.K. acknowledges support by NRF Grant funded by the Korean Government (NRF-2011-357-C00022). L. M. acknowledges support by a grant from the Swiss National Science Foundation (SNSF)(project PA00P2-136420).
\end{acknowledgments}


\begin{thebibliography}{10}%
\makeatletter
\providecommand \@ifxundefined [1]{%
 \ifx #1\undefined \expandafter \@firstoftwo
 \else \expandafter \@secondoftwo
\fi
}%
\providecommand \@ifnum [1]{%
 \ifnum #1\expandafter \@firstoftwo
 \else \expandafter \@secondoftwo
\fi
}%
\providecommand \enquote [1]{``#1''}%
\providecommand \bibnamefont  [1]{#1}%
\providecommand \bibfnamefont [1]{#1}%
\providecommand \citenamefont [1]{#1}%
\providecommand\href[0]{\@sanitize\@href}%
\providecommand\@href[1]{\endgroup\@@startlink{#1}\endgroup\@@href}%
\providecommand\@@href[1]{#1\@@endlink}%
\providecommand \@sanitize [0]{\begingroup\catcode`\&12\catcode`\#12\relax}%
\@ifxundefined \pdfoutput {\@firstoftwo}{%
 \@ifnum{\z@=\pdfoutput}{\@firstoftwo}{\@secondoftwo}%
}{%
 \providecommand\@@startlink[1]{\leavevmode}%
 \providecommand\@@endlink[0]{}%
}{%
 \providecommand\@@startlink[1]{%
  \leavevmode
  \pdfstartlink
   attr{/Border[0 0 1 ]/H/I/C[0 1 1]}%
   user{/Subtype/Link/A<</Type/Action/S/URI/URI(#1)>>}%
  \relax
 }%
 \providecommand\@@endlink[0]{\pdfendlink}%
}%
\providecommand \url  [0]{\begingroup\@sanitize \@url }%
\providecommand \@url [1]{\endgroup\@href {#1}{\urlprefix}}%
\providecommand \urlprefix [0]{URL }%
\providecommand \Eprint[0]{\href }%
\@ifxundefined \urlstyle {%
  \providecommand \doi [1]{doi:\discretionary{}{}{}#1}%
}{%
  \providecommand \doi [0]{doi:\discretionary{}{}{}\begingroup
  \urlstyle{rm}\Url }%
}%
\providecommand \doibase [0]{http://dx.doi.org/}%
\providecommand \Doi[1]{\href{\doibase#1}}%
\providecommand \bibAnnote [3]{%
  \BibitemShut{#1}%
  \begin{quotation}\noindent
    \textsc{Key:}\ #2\\\textsc{Annotation:}\ #3%
  \end{quotation}%
}%
\providecommand \bibAnnoteFile [2]{%
  \IfFileExists{#2}{\bibAnnote {#1} {#2} {\input{#2}}}{}%
}%
\providecommand \typeout [0]{\immediate \write \m@ne }%
\providecommand \selectlanguage [0]{\@gobble}%
\providecommand \bibinfo [0]{\@secondoftwo}%
\providecommand \bibfield [0]{\@secondoftwo}%
\providecommand \translation [1]{[#1]}%
\providecommand \BibitemOpen[0]{}%
\providecommand \bibitemStop [0]{}%
\providecommand \bibitemNoStop [0]{.\EOS\space}%
\providecommand \EOS [0]{\spacefactor3000\relax}%
\providecommand \BibitemShut [1]{\csname bibitem#1\endcsname}%
\bibitem{Geim:2007p4020}%
  \BibitemOpen
  \bibfield{author}{%
  \bibinfo {author} {\bibfnamefont{A.~K.}\ \bibnamefont{Geim}}\ and\ \bibinfo
  {author} {\bibfnamefont{K.~S.}\ \bibnamefont{Novoselov}},\ }%
  \bibfield{journal}{%
  \bibinfo {journal} {Nat. Mat.}\ }%
  \textbf{\bibinfo {volume} {6}},\ \bibinfo {pages} {183} (\bibinfo {month}
  {Mar.}\ \bibinfo {year} {2007})%
  \bibAnnoteFile{NoStop}{Geim:2007p4020}%
\bibitem{Berger:2004jh}%
  \BibitemOpen
  \bibfield{author}{%
  \bibinfo {author} {\bibfnamefont{C.}~\bibnamefont{Berger}}, \bibinfo {author}
  {\bibfnamefont{Z.}~\bibnamefont{Song}}, \bibinfo {author}
  {\bibfnamefont{T.}~\bibnamefont{Li}}, \bibinfo {author}
  {\bibfnamefont{X.}~\bibnamefont{Li}}, \bibinfo {author}
  {\bibfnamefont{A.~Y.}\ \bibnamefont{Ogbazghi}}, \bibinfo {author}
  {\bibfnamefont{R.}~\bibnamefont{Feng}}, \bibinfo {author}
  {\bibfnamefont{Z.}~\bibnamefont{Dai}}, \bibinfo {author}
  {\bibfnamefont{A.~N.}\ \bibnamefont{Marchenkov}}, \bibinfo {author}
  {\bibfnamefont{E.~H.}\ \bibnamefont{Conrad}}, \bibinfo {author}
  {\bibfnamefont{P.~N.}\ \bibnamefont{First}},\ and\ \bibinfo {author}
  {\bibfnamefont{W.~A.}\ \bibnamefont{de~Heer}},\ }%
  \bibfield{journal}{%
  \bibinfo {journal} {J. Phys. Chem. B}\ }%
  \textbf{\bibinfo {volume} {108}},\ \bibinfo {pages} {19912} (\bibinfo {month}
  {Dec.}\ \bibinfo {year} {2004})%
  \bibAnnoteFile{NoStop}{Berger:2004jh}%
\bibitem{Emtsev:2009p212}%
  \BibitemOpen
  \bibfield{author}{%
  \bibinfo {author} {\bibfnamefont{K.~V.}\ \bibnamefont{Emtsev}}, \bibinfo
  {author} {\bibfnamefont{A.}~\bibnamefont{Bostwick}}, \bibinfo {author}
  {\bibfnamefont{K.}~\bibnamefont{Horn}}, \bibinfo {author}
  {\bibfnamefont{J.}~\bibnamefont{Jobst}}, \bibinfo {author}
  {\bibfnamefont{G.~L.}\ \bibnamefont{Kellogg}}, \bibinfo {author}
  {\bibfnamefont{L.}~\bibnamefont{Ley}}, \bibinfo {author}
  {\bibfnamefont{J.~L.}\ \bibnamefont{McChesney}}, \bibinfo {author}
  {\bibfnamefont{T.}~\bibnamefont{Ohta}}, \bibinfo {author}
  {\bibfnamefont{S.~A.}\ \bibnamefont{Reshanov}}, \bibinfo {author}
  {\bibfnamefont{J.}~\bibnamefont{Rohrl}}, \bibinfo {author}
  {\bibfnamefont{E.}~\bibnamefont{Rotenberg}}, \bibinfo {author}
  {\bibfnamefont{A.~K.}\ \bibnamefont{Schmid}}, \bibinfo {author}
  {\bibfnamefont{D.}~\bibnamefont{Waldmann}}, \bibinfo {author}
  {\bibfnamefont{H.~B.}\ \bibnamefont{Weber}},\ and\ \bibinfo {author}
  {\bibfnamefont{T.}~\bibnamefont{Seyller}},\ }%
  \bibfield{journal}{%
  \bibinfo {journal} {Nat. Mat.}\ }%
  \textbf{\bibinfo {volume} {8}},\ \bibinfo {pages} {203} (\bibinfo {month}
  {Feb.}\ \bibinfo {year} {2009})%
  \bibAnnoteFile{NoStop}{Emtsev:2009p212}%
\bibitem{Emtsev:2008p4975}%
  \BibitemOpen
  \bibfield{author}{%
  \bibinfo {author} {\bibfnamefont{K.}~\bibnamefont{Emtsev}}, \bibinfo {author}
  {\bibfnamefont{F.}~\bibnamefont{Speck}}, \bibinfo {author}
  {\bibfnamefont{T.}~\bibnamefont{Seyller}}, \bibinfo {author}
  {\bibfnamefont{L.}~\bibnamefont{Ley}},\ and\ \bibinfo {author}
  {\bibfnamefont{J.}~\bibnamefont{Riley}},\ }%
  \bibfield{journal}{%
  \bibinfo {journal} {Phys. Rev. B}\ }%
  \textbf{\bibinfo {volume} {77}},\ \bibinfo {pages} {155303} (\bibinfo {month}
  {Apr.}\ \bibinfo {year} {2008})%
  \bibAnnoteFile{NoStop}{Emtsev:2008p4975}%
\bibitem{MuchaKruczynski:2008p214}%
  \BibitemOpen
  \bibfield{author}{%
  \bibinfo {author} {\bibfnamefont{M.}~\bibnamefont{Mucha-Kruczy{\'n}ski}},
  \bibinfo {author} {\bibfnamefont{O.}~\bibnamefont{Tsyplyatyev}}, \bibinfo
  {author} {\bibfnamefont{A.}~\bibnamefont{Grishin}}, \bibinfo {author}
  {\bibfnamefont{E.}~\bibnamefont{McCann}}, \bibinfo {author}
  {\bibfnamefont{V.}~\bibnamefont{Fal'ko}}, \bibinfo {author}
  {\bibfnamefont{A.}~\bibnamefont{Bostwick}},\ and\ \bibinfo {author}
  {\bibfnamefont{E.}~\bibnamefont{Rotenberg}},\ }%
  \bibfield{journal}{%
  \bibinfo {journal} {Phys. Rev. B}\ }%
  \textbf{\bibinfo {volume} {77}},\ \bibinfo {pages} {195403} (\bibinfo {month}
  {May}\ \bibinfo {year} {2008})%
  \bibAnnoteFile{NoStop}{MuchaKruczynski:2008p214}%
\bibitem{McChesney:2008p261}%
  \BibitemOpen
  \bibfield{author}{%
  \bibinfo {author} {\bibfnamefont{J.}~\bibnamefont{McChesney}}, \bibinfo
  {author} {\bibfnamefont{A.}~\bibnamefont{Bostwick}}, \bibinfo {author}
  {\bibfnamefont{T.}~\bibnamefont{Ohta}},\ and\ \bibinfo {author}
  {\bibfnamefont{K.}~\bibnamefont{Emtsev}},\ }%
  \bibfield{journal}{%
  \bibinfo {journal} {arXiv},\ \bibinfo {pages} {0809.4046v1}}%
   (\bibinfo {month} {Sep.}\ \bibinfo {year} {2008})%
  \bibAnnoteFile{NoStop}{McChesney:2008p261}%
\bibitem{Polini:2008p3589}%
  \BibitemOpen
  \bibfield{author}{%
  \bibinfo {author} {\bibfnamefont{M.}~\bibnamefont{Polini}}, \bibinfo {author}
  {\bibfnamefont{R.}~\bibnamefont{Asgari}}, \bibinfo {author}
  {\bibfnamefont{G.}~\bibnamefont{Borghi}}, \bibinfo {author}
  {\bibfnamefont{Y.}~\bibnamefont{Barlas}},\ and\ \bibinfo {author}
  {\bibfnamefont{T.}~\bibnamefont{Pereg-Barnea}},\ }%
  \bibfield{journal}{%
  \bibinfo {journal} {Phys. Rev. B}}%
   (\bibinfo {month} {Jan.}\ \bibinfo {year} {2008})%
  \bibAnnoteFile{NoStop}{Polini:2008p3589}%
\bibitem{Hwang:2008p3441}%
  \BibitemOpen
  \bibfield{author}{%
  \bibinfo {author} {\bibfnamefont{E.}~\bibnamefont{Hwang}}\ and\ \bibinfo
  {author} {\bibfnamefont{S.}~\bibnamefont{das Sarma}},\ }%
  \bibfield{journal}{%
  \bibinfo {journal} {Phys. Rev. B}\ }%
  \textbf{\bibinfo {volume} {77}},\ \bibinfo {pages} {081412} (\bibinfo {month}
  {Feb.}\ \bibinfo {year} {2008})%
  \bibAnnoteFile{NoStop}{Hwang:2008p3441}%
\bibitem{Bostwick:2010p3387}%
  \BibitemOpen
  \bibfield{author}{%
  \bibinfo {author} {\bibfnamefont{A.}~\bibnamefont{Bostwick}}, \bibinfo
  {author} {\bibfnamefont{F.}~\bibnamefont{Speck}}, \bibinfo {author}
  {\bibfnamefont{T.}~\bibnamefont{Seyller}}, \bibinfo {author}
  {\bibfnamefont{K.}~\bibnamefont{Horn}}, \bibinfo {author}
  {\bibfnamefont{M.}~\bibnamefont{Polini}}, \bibinfo {author}
  {\bibfnamefont{R.}~\bibnamefont{Asgari}}, \bibinfo {author}
  {\bibfnamefont{A.~H.}\ \bibnamefont{MacDonald}},\ and\ \bibinfo {author}
  {\bibfnamefont{E.}~\bibnamefont{Rotenberg}},\ }%
  \bibfield{journal}{%
  \bibinfo {journal} {Science}\ }%
  \textbf{\bibinfo {volume} {328}},\ \bibinfo {pages} {999} (\bibinfo {month}
  {May}\ \bibinfo {year} {2010})%
  \bibAnnoteFile{NoStop}{Bostwick:2010p3387}%
\bibitem{Qi:2010p4916}%
  \BibitemOpen
  \bibfield{author}{%
  \bibinfo {author} {\bibfnamefont{Y.}~\bibnamefont{Qi}}, \bibinfo {author}
  {\bibfnamefont{S.}~\bibnamefont{Rhim}}, \bibinfo {author}
  {\bibfnamefont{G.}~\bibnamefont{Sun}}, \bibinfo {author}
  {\bibfnamefont{M.}~\bibnamefont{Weinert}},\ and\ \bibinfo {author}
  {\bibfnamefont{L.}~\bibnamefont{Li}},\ }%
  \bibfield{journal}{%
  \bibinfo {journal} {Phys. Rev. Lett.}\ }%
  \textbf{\bibinfo {volume} {105}},\ \bibinfo {pages} {085502} (\bibinfo
  {month} {Aug.}\ \bibinfo {year} {2010})%
  \bibAnnoteFile{NoStop}{Qi:2010p4916}%
\bibitem{Gierz:2011do}%
  \BibitemOpen
  \bibfield{author}{%
  \bibinfo {author} {\bibfnamefont{I.}~\bibnamefont{Gierz}}, \bibinfo {author}
  {\bibfnamefont{J.}~\bibnamefont{Henk}}, \bibinfo {author}
  {\bibfnamefont{H.}~\bibnamefont{H{\"o}chst}}, \bibinfo {author}
  {\bibfnamefont{C.~R.}\ \bibnamefont{Ast}},\ and\ \bibinfo {author}
  {\bibfnamefont{K.}~\bibnamefont{Kern}},\ }%
  \bibfield{journal}{%
  \bibinfo {journal} {Phys. Rev. B}\ }%
  \textbf{\bibinfo {volume} {83}},\ \bibinfo {pages} {121408(R)} (\bibinfo
  {month} {Mar.}\ \bibinfo {year} {2011})%
  \bibAnnoteFile{NoStop}{Gierz:2011do}%
\bibitem{Siegel:2011va}%
  \BibitemOpen
  \bibfield{author}{%
  \bibinfo {author} {\bibfnamefont{D.~A.}\ \bibnamefont{Siegel}}, \bibinfo
  {author} {\bibfnamefont{C.-H.}\ \bibnamefont{Park}}, \bibinfo {author}
  {\bibfnamefont{C.}~\bibnamefont{Hwang}}, \bibinfo {author}
  {\bibfnamefont{J.}~\bibnamefont{Deslippe}}, \bibinfo {author}
  {\bibfnamefont{A.~V.}\ \bibnamefont{Fedorov}}, \bibinfo {author}
  {\bibfnamefont{S.~G.}\ \bibnamefont{Louie}},\ and\ \bibinfo {author}
  {\bibfnamefont{A.}~\bibnamefont{Lanzara}},\ }%
  \bibfield{journal}{%
  \bibinfo {journal} {Proceedings of the National Academy of Sciences}\ }%
  \textbf{\bibinfo {volume} {108}} (\bibinfo {month} {Jun.}\ \bibinfo {year}
  {2011})%
  \bibAnnoteFile{NoStop}{Siegel:2011va}%
\bibitem{Kim:2008p4920}%
  \BibitemOpen
  \bibfield{author}{%
  \bibinfo {author} {\bibfnamefont{S.}~\bibnamefont{Kim}}, \bibinfo {author}
  {\bibfnamefont{J.}~\bibnamefont{Ihm}}, \bibinfo {author}
  {\bibfnamefont{H.}~\bibnamefont{Choi}},\ and\ \bibinfo {author}
  {\bibfnamefont{Y.-W.}\ \bibnamefont{Son}},\ }%
  \bibfield{journal}{%
  \bibinfo {journal} {Phys. Rev. Lett.}\ }%
  \textbf{\bibinfo {volume} {100}},\ \bibinfo {pages} {176802} (\bibinfo
  {month} {Apr.}\ \bibinfo {year} {2008})%
  \bibAnnoteFile{NoStop}{Kim:2008p4920}%
\bibitem{Zhou:2007p4898}%
  \BibitemOpen
  \bibfield{author}{%
  \bibinfo {author} {\bibfnamefont{S.~Y.}\ \bibnamefont{Zhou}}, \bibinfo
  {author} {\bibfnamefont{G.}~\bibnamefont{Gweon}}, \bibinfo {author}
  {\bibfnamefont{A.~V.}\ \bibnamefont{Fedorov}}, \bibinfo {author}
  {\bibfnamefont{P.~N.}\ \bibnamefont{First}}, \bibinfo {author}
  {\bibfnamefont{W.~A.}\ \bibnamefont{DeHeer}}, \bibinfo {author}
  {\bibfnamefont{D.}~\bibnamefont{Lee}}, \bibinfo {author}
  {\bibfnamefont{F.}~\bibnamefont{Guinea}}, \bibinfo {author}
  {\bibfnamefont{A.}~\bibnamefont{Neto}},\ and\ \bibinfo {author}
  {\bibfnamefont{A.}~\bibnamefont{Lanzara}},\ }%
  \bibfield{journal}{%
  \bibinfo {journal} {Nat. Mat.}\ }%
  \textbf{\bibinfo {volume} {6}},\ \bibinfo {pages} {770} (\bibinfo {month}
  {Sep.}\ \bibinfo {year} {2007})%
  \bibAnnoteFile{NoStop}{Zhou:2007p4898}%
\bibitem{Bostwick:2009p256}%
  \BibitemOpen
  \bibfield{author}{%
  \bibinfo {author} {\bibfnamefont{A.}~\bibnamefont{Bostwick}}, \bibinfo
  {author} {\bibfnamefont{J.}~\bibnamefont{McChesney}}, \bibinfo {author}
  {\bibfnamefont{K.}~\bibnamefont{Emtsev}},\ and\ \bibinfo {author}
  {\bibfnamefont{T.}~\bibnamefont{Seyller}},\ }%
  \bibfield{journal}{%
  \bibinfo {journal} {Phys. Rev. Lett.}}%
   (\bibinfo {month} {Jan.}\ \bibinfo {year} {2009})%
  \bibAnnoteFile{NoStop}{Bostwick:2009p256}%
\bibitem{deGail:2011cl}%
  \BibitemOpen
  \bibfield{author}{%
  \bibinfo {author} {\bibfnamefont{R.}~\bibnamefont{de~Gail}}, \bibinfo
  {author} {\bibfnamefont{M.}~\bibnamefont{Goerbig}}, \bibinfo {author}
  {\bibfnamefont{F.}~\bibnamefont{Guinea}}, \bibinfo {author}
  {\bibfnamefont{G.}~\bibnamefont{Montambaux}},\ and\ \bibinfo {author}
  {\bibfnamefont{A.}~\bibnamefont{Castro~Neto}},\ }%
  \bibfield{journal}{%
  \bibinfo {journal} {Phys. Rev. B}\ }%
  \textbf{\bibinfo {volume} {84}} (\bibinfo {month} {Jul.}\ \bibinfo {year}
  {2011})%
  \bibAnnoteFile{NoStop}{deGail:2011cl}%
\bibitem{Bostwick:2007p252}%
  \BibitemOpen
  \bibfield{author}{%
  \bibinfo {author} {\bibfnamefont{A.}~\bibnamefont{Bostwick}}, \bibinfo
  {author} {\bibfnamefont{T.}~\bibnamefont{Ohta}}, \bibinfo {author}
  {\bibfnamefont{J.}~\bibnamefont{McChesney}}, \bibinfo {author}
  {\bibfnamefont{T.}~\bibnamefont{Seyller}},\ and\ \bibinfo {author}
  {\bibfnamefont{E.}~\bibnamefont{Rotenberg}},\ }%
  \bibfield{journal}{%
  \bibinfo {journal} {Solid State Comm.}\ }%
  \textbf{\bibinfo {volume} {143}},\ \bibinfo {pages} {63} (\bibinfo {month}
  {Jan.}\ \bibinfo {year} {2007})%
  \bibAnnoteFile{NoStop}{Bostwick:2007p252}%
\bibitem{Ostler:2010p5375}%
  \BibitemOpen
  \bibfield{author}{%
  \bibinfo {author} {\bibfnamefont{M.}~\bibnamefont{Ostler}}, \bibinfo {author}
  {\bibfnamefont{F.}~\bibnamefont{Speck}}, \bibinfo {author}
  {\bibfnamefont{M.}~\bibnamefont{Gick}},\ and\ \bibinfo {author}
  {\bibfnamefont{T.}~\bibnamefont{Seyller}},\ }%
  \bibfield{journal}{%
  \bibinfo {journal} {Phys. Stat. Sol. B}\ }%
  \textbf{\bibinfo {volume} {247}},\ \bibinfo {pages} {2924} (\bibinfo {month}
  {Sep.}\ \bibinfo {year} {2010})%
  \bibAnnoteFile{NoStop}{Ostler:2010p5375}%
\bibitem{Bostwick:2009p255}%
  \BibitemOpen
  \bibfield{author}{%
  \bibinfo {author} {\bibfnamefont{A.}~\bibnamefont{Bostwick}}, \bibinfo
  {author} {\bibfnamefont{J.}~\bibnamefont{McChesney}}, \bibinfo {author}
  {\bibfnamefont{T.}~\bibnamefont{Ohta}},\ and\ \bibinfo {author}
  {\bibfnamefont{E.}~\bibnamefont{Rotenberg}},\ }%
  \bibfield{journal}{%
  \bibinfo {journal} {Prog. Surf. Sci.}}%
   (\bibinfo {month} {Jan.}\ \bibinfo {year} {2009})%
  \bibAnnoteFile{NoStop}{Bostwick:2009p255}%
\bibitem{Bostwick:2007p247}%
  \BibitemOpen
  \bibfield{author}{%
  \bibinfo {author} {\bibfnamefont{A.}~\bibnamefont{Bostwick}}, \bibinfo
  {author} {\bibfnamefont{T.}~\bibnamefont{Ohta}}, \bibinfo {author}
  {\bibfnamefont{T.}~\bibnamefont{Seyller}}, \bibinfo {author}
  {\bibfnamefont{K.}~\bibnamefont{Horn}},\ and\ \bibinfo {author}
  {\bibfnamefont{E.}~\bibnamefont{Rotenberg}},\ }%
  \bibfield{journal}{%
  \bibinfo {journal} {Nat. Phys.}\ }%
  \textbf{\bibinfo {volume} {3}},\ \bibinfo {pages} {36} (\bibinfo {month}
  {Dec.}\ \bibinfo {year} {2006})%
  \bibAnnoteFile{NoStop}{Bostwick:2007p247}%
\bibitem{Walter:2011uu}%
  \BibitemOpen
  \bibfield{author}{%
  \bibinfo {author} {\bibfnamefont{A.~L.}\ \bibnamefont{Walter}}, \bibinfo
  {author} {\bibfnamefont{S.}~\bibnamefont{Nie}}, \bibinfo {author}
  {\bibfnamefont{A.}~\bibnamefont{Bostwick}}, \bibinfo {author}
  {\bibfnamefont{k.~s.}\ \bibnamefont{kim}}, \bibinfo {author}
  {\bibfnamefont{L.}~\bibnamefont{Moreschini}}, \bibinfo {author}
  {\bibfnamefont{Y.~J.}\ \bibnamefont{Chang}}, \bibinfo {author}
  {\bibfnamefont{D.}~\bibnamefont{Innocenti}}, \bibinfo {author}
  {\bibfnamefont{K.}~\bibnamefont{Horn}}, \bibinfo {author}
  {\bibfnamefont{K.~F.}\ \bibnamefont{McCarty}},\ and\ \bibinfo {author}
  {\bibfnamefont{E.}~\bibnamefont{Rotenberg}},\ }%
  \bibfield{journal}{%
  \bibinfo {journal} {Phys. Rev. B}\ }%
  \textbf{\bibinfo {volume} {84}} (\bibinfo {month} {Sep.}\ \bibinfo {year}
  {2011})%
  \bibAnnoteFile{NoStop}{Walter:2011uu}%
\bibitem{Sutter:2009ff}%
  \BibitemOpen
  \bibfield{author}{%
  \bibinfo {author} {\bibfnamefont{P.}~\bibnamefont{Sutter}}, \bibinfo {author}
  {\bibfnamefont{J.~T.}\ \bibnamefont{Sadowski}},\ and\ \bibinfo {author}
  {\bibfnamefont{E.}~\bibnamefont{Sutter}},\ }%
  \bibfield{journal}{%
  \bibinfo {journal} {Phys. Rev. B}\ }%
  \textbf{\bibinfo {volume} {80}} (\bibinfo {month} {Dec.}\ \bibinfo {year}
  {2009})%
  \bibAnnoteFile{NoStop}{Sutter:2009ff}%
\end{thebibliography}

%
\end{document}